\documentclass[12pt]{article}
\usepackage{amsfonts}

\usepackage{pstricks}
\usepackage{pst-node}
\usepackage{epsfig}
\usepackage{amsmath,amssymb,latexsym}
\usepackage{xcolor}
\usepackage{hyperref}
\usepackage{slashed} 

\usepackage{import}

\usepackage{enumerate}

\usepackage{physics}

\usepackage[lmargin=1.0in, rmargin=1.0in, tmargin=1.0in, bmargin=1.0in]{geometry}

\usepackage{cancel}

\usepackage{empheq}

\usepackage{comment}

\usepackage{setspace}
\begin{document}



\title{\bf Reformulation of continuum defects in terms of the general teleparallel geometry in the language of exterior algebra}
\author{  Muzaffer Adak$^1$, Tekin Dereli$^{2,3}$, Ertan Kok$^1$, Ozcan Sert$^1$  \\
  {\small $^1$Department of Physics, Faculty of Science, Pamukkale University, Denizli, Türkiye} \\
  {\small $^2$Department of Basic Sciences
Faculty of Engineering and Natural Sciences,} \\
{\small Maltepe University, 34857 Maltepe, Istanbul, T\"urkiye}\\
  {\small $^3$Department of Physics, Faculty of Science, Ko\c{c} University, 34450 Sariyer, Istanbul, T\"urkiye} \\
      {\small {\it E-mail:} {\blue madak@pau.edu.tr, tdereli@ku.edu.tr, ekok@pau.edu.tr, osert@pau.edu.tr}}}

  \vskip 1cm
\date{\today}
\maketitle
\thispagestyle{empty}
\begin{abstract}
 \noindent
We discuss the geometric formulation of continuum defects consisting of dislocations and disclinations. After reviewing the metric affine geometry and the present geometric formulation of dislocation and disclination written in terms of torsion and full curvature (together with vanishing non-metricity), we give a new formulation of them in a novel way in terms of torsion and non-metricity (together with vanishing full curvature), the so-called general teleparallel geometry. All calculations are performed by using the exterior algebra. We obtain continuity equations explicitly for dislocation density and disclination density. \\



\end{abstract}

\section{Introduction}

It is known that the role of geometry in physics is very valuable. The most familiar example of it is general relativity, Einstein's theory of gravity, in which the gravitational field is associated with the Riemann curvature tensor. With similar motivations, geometric quantities have been used to express the theory of defects in materials. In this context, the physical state of a material describes the geometry of the space occupied by its atoms. In studies conducted for this purpose so far, dislocation type defects have been paired with torsion and disclination type defects with full curvature \cite{rivier1990}-\cite{katanaev2021}. We review main results of those works in Section \ref{sec:defect-theory-present} in terms of exterior forms. There are works discussing the point defects apart from these two line defects as well and their possible relations to the non-metricity \cite{gunther1985}-\cite{yavari-2012}. In addition to these works, the authors of Ref \cite{Volovik-1980} investigate the nonlinear equations of the elasticity theory for bodies with impurities, dislocations and disclinations by adhering a novel mathematical perspective, more specifically by using the method of the Poisson “hydrodynamic” brackets. The papers \cite{anthony1970},\cite{anthony1971} may be seen as the first works discussing the possible relations among the non-metric connections (i.e. non-metricity), quasi-dislocations and quasi-disclinations.

In this work we adopt the terminology used by de Wit in the papers \cite{dewit1970i}-\cite{dewit1973iv} published on the theory of continuum defects. Accordingly, in 1907, Volterra studied the elastic fields for a long straight isotropic cylinder containing dislocations and disclinations \cite{voltera1907} and named these defects as distortions. In 1920, Love called them as dislocations \cite{love1920}. Love's nomenclature has become widely accepted, but its meaning has become restricted to only translational dislocations by suppressing rotational dislocations. Therefore, in 1958 Frank introduced a new name, disclination (the first version was disinclination), to describe rotational dislocation \cite{frank1958}. Since then, the word ``distortion" has been used for the gradient of displacement. Correspondingly, Volterra's original word ``distortion" (dislocation and disclination) is sometimes confused with today's word ``distortion" (gradient of displacement). To overcome this confusion, it is prefered to use the term ``defect" (dislocation and disclination) instead of Volterra's original word ``distortion". The word ``defects" in this terminology may seem to refer to line defects in crystal researches. However, the usage of the word ``defect" does not exclude point defects in crystal studies, because a point defect can be considered as a defect loop (closed line defect). In summary, disclination is a defect in the orientation of director whereas a dislocation is a defect in positional order.

Dislocation density $\alpha_{ab}$ and disclination density $\theta_{ab}$ are very convenient tools to obtain the most useful elastic fields such as distortion, strain, dilatation, stress and rotation caused by the presence of dislocations and disclinations in a continuum. Physically, as the diagonal components of dislocation density represents the screw type translational defects, the off-diagonal ones are related with the edge type translational defects. Analogously, as the diagonal components of disclination density are associated with the wedge type of rotational defects, the off-diagonal components measure the twist type rotational defects \cite{dewit1973iii}. In the absence of disclination, the dislocation density is defined self-consistently and is directly related to the Burgers vector $\Vec{b}$ via the formula $b^a := \iint_S \alpha^{ba} *e_b$ where the term $*e_b$ represents the oriented area element on the surface $S$. This relation means that the dislocation density $\alpha^{ba}$ represents the flux of Burgers vectors in the $x_a$ direction that pierces unit area of the surface $S$ normal to the $x_b$ direction. Moreover, in the presence of disclination, it shifts the dislocation density and therefore in addition to the (general) Burgers vector $\Vec{B}$  a new vector, the so-called the Frank vector $\Vec{\Omega}$, could be defined corresponding to the disclination density. Their definitions given by \cite{dewit1973ii} will be recast in the Section \ref{sec:defect-theory-present}.

However, if torsion is paired with dislocation and full curvature with disclination as in the geometric picture we summarized above, there will be no shift in dislocation when disclination occurs. On the contrary, the presence of dislocation causes a shift in disclination as shown in the papers \cite{dewit1970i}-\cite{dewit1973iv}. This contradiction is the first motivation leading us to do this work.

On the other hand, we know that a metric affine geometry is defined by the triple; manifold, metric, affine connection or correspondingly by the triple; non-metricity, torsion, curvature. Furthermore, the affine connection 1-form, $\omega_{ab}$, contains a Riemannian component (the so-called Levi-Civita connection 1-form, $\widetilde{\omega}_{ab}$, computed from metric) and a non-Riemannian component (which we call as defect tensor-valued 1-form, $\mathrm{L}_{ab}$, consisting of tensors of torsion and non-metricity). Although in the literature of geometry and gravity it is usually called as distortion 1-form, we prefer to name it as defect 1-form appropriate to the topic. We summarize the basics of metric affine geometry in the forthcoming section. More specifically, while the symmetric part of defect 1-form consist of only non-metricity 1-form, its anti-symmetric part contains torsion and certain components of non-metricity. Thus, when the non-metricity vanishes, the defect 1-form turns out to be anti-symmetric and is given only by torsion. However, when non-metricity does not vanish, along with that the symmetric part of the defect 1-form appears, its anti-symmetric part is shifted by certain components of non-metricity. One can consult Ref.\cite{adak2023ijgmmp} for further reading. In conclusion, the resemblances between the statements on dislocation (disclination) and the anti-symmetric part (symmetric part) of defect 1-form led us to investigate the reformulation of the defect theory of a continuum in terms of Cartan-Weyl (general teleparallel) geometry. Besides all these, on the geometry side, the general teleparallel geometry has an advantage over the Riemann-Cartan geometry. While the latter does not have a metric formulation, the former does. Metric formulation of general teleparallel geometry is the second motivation leading us to do this work. In fact, all three teleparallel geometries have  metric formulations \cite{adak2023ijgmmp}. As far as we know, our general teleparallel geometric formulation of continuum defects is the first attempt in the literature. Free from special discussion here, although plenty of papers has been published, no fundamental theory of defects is yet available. We develop our formulation for any non-Cartesian coordinates covariantly in the language of exterior algebra.

The plan of our paper is as follows. Since our work is based on mainly the geometry and the exterior algebra, we summarize our mathematical preliminaries in the next section. Then we review the present literature on geometric formulation of the theory of continuum defects in Section \ref{sec:defect-theory-present}.
Later, we give our novel theory of continuum defects written in the general teleparallel geometry in Section \ref{sec:new-theory}.
In the section of Discussion, we collect our results and state our future projects as continuation of this paper. Besides, since the continuity equations of dislocation density and disclination density are derived mainly through the Bianchi identities relating non-metricity, torsion and full curvature to each other, we give some detail on them in the language of exterior algebra in the section of Appendix.

\section{Metric affine geometry}

The triple $\{M,g,\nabla\}$ or $\{Q_{ab},T^a,R^a{}_b\}$ defines a metric affine geometry where $M$ is three-dimensional orientable and differentiable manifold, $g$ is non-degenerate symmetric metric, $\nabla$ is full (or affine) connection or covariant derivative, $Q_{ab}$ is non-metricity 1-form, $T^a$ is torsion 2-form and $R^a{}_b$ is curvature 2-form \cite{thirring1997}, \cite{frankel2012}. We denote the anholonomic metric-orthonormal coframe by $e^a$, then write the metric as $g=g_{ab} e^a \otimes e^b$ where $g_{ab}:=\delta_{ab}$ is the Euclid metric with $\delta_{ab}=\text{diag}(1,1,1)$. The full connection is determined by the full connection 1-form $\omega^a{}_b$ via the definition $\nabla e^a := - \omega^a{}_b \wedge e^b$ or $\nabla X_a := X_b \otimes \omega^b{}_a$ where $\otimes$ denotes tensor product, $\wedge$ is the exterior product in the exterior algebra and $X_a$ is the orthonormal frame such that $e^a(X_b)=\delta^a_b$ is the duality relation. Besides, $e^a$ is called the metric orthonormal (or shortly orthonormal) 1-form and the Cartan structure equations are given by tensor-valued non-metricity 1-form\footnote{Here $\omega_{(ab)} = Q_{ab}$ is the reason why we adopt the factor $-\frac{1}{2}$ in the definition of non-metricity}, tensor-valued torsion 2-form and tensor-valued full (or non-Riemannian) curvature 2-form, respectively,
 \begin{subequations}\label{eq:cartan-ort}
 \begin{align}
     Q_{ab} &:= -\frac{1}{2} D\delta_{ab}  = -\frac{1}{2} \left( d\delta_{ab} - \omega^c{}_a \delta_{cb} - \omega^c{}_b \delta_{ac} \right) = \omega_{(ab)} , \label{eq:nonmetric}\\
     T^a &:= De^a = de^a + \omega^a{}_b \wedge e^b, \label{eq:tors}\\
     R^a{}_b &:= D\omega^a{}_b := d \omega^a{}_b + \omega^a{}_c \wedge \omega^c{}_b, \label{eq:curv}
 \end{align}
 \end{subequations}
where $d$ is the exterior derivative and $D$ is the covariant exterior derivative. We used the result $d\delta_{ab}=0$ in the orthonormal frame in (\ref{eq:nonmetric}). We utilize the usual notation: $( \bullet )$ and $[ \bullet ]$ means to calculate the symmetric and anti-symmetric parts with respect to the indices enclosed, i.e., $  \omega_{(ab)} = \frac{1}{2} (\omega_{ab} + \omega_{ba})$ and $  \omega_{[ab]} = \frac{1}{2} (\omega_{ab} - \omega_{ba})$. They satisfy the Bianchi identities
  \begin{subequations} \label{eq:bianchi_identities}
  \begin{align}
      DQ_{ab} =& R_{(ab)} , \label{eq:bianchi1} \\
      DT^a =& R^a{}_b \wedge e^b , \label{eq:bianchi2}\\
      DR^a{}_b = &0 . \label{eq:bianchi3}
  \end{align}
  \end{subequations}
Accordingly, we define the covariant exterior derivative of any $(p,q)$-type tensor-valued exterior form $\mathfrak{T}^{a_1 a_2 \cdots a_p}_{\; \; \; \; b_1 b_2 \cdots b_q }$ below
 \begin{eqnarray}\label{eq:gl-covariant-derivative}
   D \mathfrak{T}^{a_1 a_2 \cdots a_p}_{\; \; \; \; b_1 b_2 \cdots b_q } & = d \mathfrak{T}^{a_1 a_2 \cdots a_p}_{\; \; \; \; b_1 b_2 \cdots b_q }
    + \omega^{a_1}{}_c \wedge \mathfrak{T}^{c a_2 \cdots a_p}_{\; \; \; \; b_1 b_2 \cdots b_q } + \cdots + \omega^{a_p}{}_c \wedge \mathfrak{T}^{a_1 a_2 \cdots c}_{\; \; \; \; b_1 b_2 \cdots b_q } \nonumber \\
     & \quad 
     - \omega^c{}_{b_1} \wedge \mathfrak{T}^{a_1 a_2 \cdots a_p}_{\; \; \; \; c b_2 \cdots b_q } - \cdots
    - \omega^c{}_{b_q} \wedge \mathfrak{T}^{a_1 a_2 \cdots a_p}_{\; \; \; \; b_1 b_2 \cdots c} .
 \end{eqnarray}
  
If need, one can pass to tensor notation easily by expanding the concerned exterior forms in components as follows
  \begin{align}
      Q_{ab}=Q_{abc} e^c , \qquad T^a = \frac{1}{2} T^a{}_{bc} e^b \wedge e^c , \qquad R^a{}_b = \frac{1}{2} R^a{}_{bcd} e^c \wedge e^d .
  \end{align}
Thus, we deduce that $Q_{abc}=Q_{(ab)c}$ is symmetric at the first two indices, $T^a{}_{bc}=T^a{}_{[bc]}$ is anti-symmetric at the last two indices and $R_{abcd} = R_{ab[cd]}$ is anti-symmetric at the last two indices and asymmetric $R_{abcd} = R_{[ab]cd} + R_{(ab)cd}$ at the first two indices by definition. In the subsequent sections, the definitions of the first kind trace 1-form $Q :=\delta_{ab} Q^{ab} = Q^a{}_{ab} e^b$ and the second kind trace 1-form $P :=(\iota_aQ^{ab}) e_b = Q^a{}_{ba} e^b$ of the non-metricity tensor will be used frequently where $\iota_a \equiv \iota_{X_a} $ denotes the interior derivative of the exterior algebra, $\iota_b e^a = \delta^a_b$. Furthermore, it is worthy to remember that the symmetric part of the first two indices of $R_{abcd}$  depends on existence of the non-metricity and will allow us to define metrical disclination density tensor in the next section. Furthermore, we would like to remind explicitly that $d\delta_{ab}=0$, $d\delta^{ab}=0$ and $d\delta^a_b=0$, but $D\delta_{ab} = -2Q_{ab}$, $D\delta^{ab}=+2Q^{ab}$ and $D\delta^a_b=0$. Accordingly, one must pay special attention as raising or lowering an index in front of $D$ operation in non-metric geometries.  

Meanwhile it is worthwhile to remark shortly on our choice of the orthonormal coframe. In this paper we adhere to formulate all quantities in the orthonormal frame (or equivalently in the orthonormal coframe). Of course, a non-orthogonal coframe can be chosen, for example, the coordinate coframe, $dx^\alpha$. Nevertheless, it is known that one can always pass from the orthonormal coframe to the coordinate coframe or vise versa at any stage of calculations via a linear transformation $dx^\alpha = h^\alpha{}_a e^a$ and its inverse $e^a = h^a{}_\alpha dx^\alpha$ where $h^\alpha{}_a$ is called as dreibein. Then, the affine connection 1-form transforms correspondingly as $ \omega^\alpha{}_\beta = h^\alpha{}_a  \omega^a{}_b h^b{}_\beta + h^\alpha{}_a dh^a{}_\beta$. Thus, tensor-valued exterior forms transform covariantly, i.e., $g_{\alpha \beta} = h^a{}_\alpha h^b{}_\beta \delta_{ab}$, $Q_{\alpha \beta} = h^a{}_\alpha h^b{}_\beta Q_{ab}$, $T^\alpha = h^\alpha{}_a T^a$, $R^\alpha{}_\beta = h^\alpha{}_a R^a{}_b h^b{}_\beta$. Finally we rewrite the Cartan structure equations in the coordinate coframe as
  \begin{subequations}
 \begin{align} 
     Q_{\alpha \beta} &:= -\frac{1}{2} Dg_{\alpha \beta}  = -\frac{1}{2} \left( dg_{\alpha \beta} - \omega^\gamma{}_\alpha g_{\gamma \beta} - \omega^\gamma{}_\beta g_{\alpha \gamma} \right) = -\frac{1}{2} dg_{\alpha \beta} + \omega_{(\alpha \beta)} ,\\
     T^\alpha &:= D (dx^\alpha) = d(dx^\alpha) + \omega^\alpha{}_\beta \wedge dx^\beta = \omega^\alpha{}_\beta \wedge dx^\beta , \\
     R^\alpha{}_\beta &:= D\omega^\alpha{}_\beta := d \omega^\alpha{}_\beta + \omega^\alpha{}_\gamma \wedge \omega^\gamma{}_\beta,
 \end{align}
 \end{subequations}
where $d(dx^\alpha) = d^2x^\alpha=0$ because of Poincare lemma in the exterior algebra. One of advantages of the usage of the orthonormal coframe is that the exterior derivative and the variation of metric components vanish, $dg_{ab}=0$ and $\delta g_{ab}=0$, but they do not in the non-orthonormal coframe,  $dg_{\alpha \beta} \neq 0$ and $\delta g_{\alpha \beta} \neq 0$. For further discussion on the orthonormal coframe, the coordinate coframe and the mixed coframe we direct the reader to the paper \cite{adak2023ijgmmp}.

\subsection{Decomposition of full connection}

The full connection 1-form, $\omega^a{}_b$, can be decomposed uniquely to a Riemannian piece, $\widetilde{\omega}_{ab}(g)$ determined by metric plus a non-Riemannian piece, $\mathrm{L}_{ab}(T,Q)$ determined by torsion and non-metricity \cite{adak2023ijgmmp},\cite{tucker1995},\cite{hehl1995}
 \begin{align}
     \omega_{ab}=\widetilde{\omega}_{ab} + \mathrm{L}_{ab}, \label{eq:connec-decom}
 \end{align}
where $\widetilde{\omega}_{ab}=-\widetilde{\omega}_{ba}$ is the Levi-Civita connection 1-form, 
 \begin{equation}\label{eq:Levi-Civita}
   \widetilde{\omega}_{ab} = \frac{1}{2} \left[ -\iota_a de_b + \iota_b de_a + (\iota_a \iota_b de_c) e^c \right] \qquad \text{or} \qquad \widetilde{\omega}^a{}_b \wedge e^b = -de^a
 \end{equation}
and $\mathrm{L}_{ab}$ is the tensor-valued defect 1-form\footnote{In gravity literature $\mathrm{L}_{ab}$ is called tensor-valued distortion 1-form.}, 
 \begin{equation}
     \mathrm{L}_{ab} = \underbrace{ \underbrace{ \frac{1}{2} \left[ \iota_a T_b - \iota_b T_a - (\iota_a \iota_b T_c) e^c \right]}_{contortion} + \underbrace{ ( \imath_b Q_{ac} - \imath_a Q_{bc} ) e^c + Q_{ab} }_{disformation} }_{defect}.
 \end{equation}
In the literature it is common to define the tensor-valued contortion  1-form, $K_{ab}=-K_{ba}$, in terms of torsion 2-form
 \begin{align} \label{eq:torsion-contor}
     K_{ab} = \frac{1}{2} \left[ \iota_a T_b - \iota_b T_a - (\iota_a \iota_b T_c) e^c \right] \qquad \text{or} \qquad K^a{}_b \wedge e^b = T^a .
 \end{align}
It is worthy to notice that the symmetric part of the affine connection is determined by only non-metricity, $\omega_{(ab)}= Q_{ab}$, the remainder of $\omega_{ab}$ is the anti-symmetric, $\omega_{[ab]}:=\Omega_{ab}= \widetilde{\omega}_{ab} + K_{ab} +  ( \imath_b Q_{ac} - \imath_a Q_{bc} ) e^c$. Correspondingly, it would be useful to write $\omega_{ab}= \Omega_{ab} + Q_{ab}$ in order to decompose the full curvature as the anti-symmetric piece plus symmetric piece, $R_{ab}=R_{[ab]} + R_{(ab)}$, where
  \begin{subequations}
      \begin{align}
          R_{[ab]} =& d \Omega_{ab} + \Omega_{ac} \wedge \Omega^c{}_b +  Q_{ac} \wedge Q^c{}_b , \label{eq:curv-ant-symm1} \\
           R_{(ab)} =& d Q_{ab} + \Omega_{ac} \wedge Q^c{}_b +  Q_{ac} \wedge \Omega^c{}_b . \label{eq:curv-ant-symm2}
      \end{align}
  \end{subequations}
Here we notice that vanishing of non-metricity terminates the symmetric part of the full curvature 2-form. Besides, all the Riemannian quantities will be labelled by a tilde over them in this paper. Accordingly, the full curvature 2-form can be split into Riemannian part plus non-Riemannian part
 \begin{align} \label{eq:decompos-curv}
     R^a{}_b = \widetilde{R}^a{}_b + \widetilde{D} \mathrm{L}^a{}_b + \mathrm{L}^a{}_c \wedge \mathrm{L}^c{}_b
 \end{align}
where $\widetilde{R}^a{}_b$ are the Riemannian curvature 2-form and $\widetilde{D} \mathrm{L}^a{}_b$ is the covariant exterior derivative of $\mathrm{L}^a{}_b$ with respect to the Levi-Civita connection,
 \begin{subequations}
     \begin{align}
         \widetilde{R}^a{}_b &= \widetilde{\omega}^a{}_{b} + \widetilde{\omega}^a{}_{c} \wedge \widetilde{\omega}^c{}_{b} , \\
         \widetilde{D}\mathrm{L}^a{}_b &= d\mathrm{L}^a{}_b +  \widetilde{\omega}^a{}_{c} \wedge \mathrm{L}^c{}_b - \widetilde{\omega}^c{}_{b} \wedge \mathrm{L}^a{}_c . \label{eq:tildaD}
     \end{align}
 \end{subequations}
When non-metricity is set to zero, the full connection is called metric compatible in which case the symmetric parts of full connection and full curvature vanish. If torsion is also reset along with non-metricity, the full connection is named the Levi-Civita (or Riemannian) connection. An affine geometry is classified whether non-metricity, torsion and/or full curvature vanish or not, see Table  \ref{tab:clasification of spacetimes}.
   \begin{table}[ht]
\caption{Classification of affine geometries. In literature, sometimes firstly $Q_{ab}$ is decomposed as $Q_{ab}=\slashed{Q}_{ab}+\frac{1}{3}\delta_{ab}Q$ where $\delta^{ab}Q_{ab}=Q$ and $\delta^{ab}\slashed{Q}_{ab}=0$, then the case of $\slashed{Q}_{ab}=0$ and $Q\neq 0$ is called Weyl geometry. But, here by ``Weyl geometry'' we mean $Q_{ab} \neq 0$ in general!}
 \centering
 \begin{tabular}{|c|c|c|l|}
 \hline
   $Q_{ab}$ & $T^a$ & $R^a{}_b$ & {\it Geometry Name}  \\
 \hline \hline
  $0$ & $0$ & $0$ & Minkowski  \\
  \hline
   $0$ & $ 0$ & $\neq 0$ & Riemann  \\
  \hline
  $0$ & $\neq 0$ & $ 0$ & Metric (Weitzenb\"ock) teleparallel  \\
  \hline
  $\neq 0$ & $0$ & $0$ & Symmetric teleparallel  \\
  \hline
  $0$ & $\neq 0$ & $\neq 0$ & Riemann-Cartan  \\
  \hline
  $\neq 0$ & $0$ & $\neq 0$ & Riemann-Weyl \\
  \hline
   $\neq 0$ & $\neq 0$ & $ 0$ & General teleparallel (Cartan-Weyl)  \\
  \hline
  $\neq 0$ & $\neq 0$ & $\neq 0$ & Riemann-Cartan-Weyl (Metric affine) \\
  \hline
 \end{tabular}
 \label{tab:clasification of spacetimes}
  \end{table}

\subsection{Some algebraic identities}

In the calculations we use the following abbreviations
  \begin{align}
      e^{ab\cdots} := e^a \wedge e^b \wedge \cdots , \quad
     \iota_{ab \cdots } := \iota_a \iota_b \cdots , \quad
     \partial_a := \iota_a d , \quad D_a:= \iota_a D ,
  \end{align}
and algebraic identities
  \begin{subequations}
 \begin{align} \label{eq:identities-algebraic}
     D*e_a &= -Q \wedge *e_a + *e_{ab} \wedge T^b,  \\
     D*e_{ab} &= -Q \wedge *e_{ab} + *e_{abc} \wedge T^c, \\
     D*e_{abc} &= -Q \wedge *e_{abc}, \label{eq:algeb-identity3}
 \end{align}
  \end{subequations}
and
  \begin{subequations} \label{eq:identities-algebraic2}
    \begin{align}
     \Lambda \wedge *\Gamma = (-1)^{pq} \Gamma \wedge * \Lambda , & \quad 
     \iota_a * \Lambda = *(\Lambda \wedge e_a) , \quad 
     e^a \wedge \iota_a \Lambda = p \Lambda , \\
     e_a \wedge *e_b = \delta_{ab} *1 , & \quad e_a \wedge *(e_b \wedge e_c) = -\delta_{ab} *e_c + \delta_{ac} *e_b , \\
     e^a \wedge e^b \wedge e^c = \epsilon^{abc}*1 , & \quad *e_a \wedge *(e_b \wedge e_c) = \epsilon_{abc} *1
     \end{align}
  \end{subequations}
where $*$ denotes the Hodge duality map, $\Lambda$ is any $p$-form, $\Gamma$ is any $q$-form and $pq$ is the ordinary multiplication of numbers of $p$ and $q$. The operation of Hodge map may be performed via the totally anti-symmetric epsilon tensor by relations
      \begin{align} \label{eq:hodge_operations}
        *1 = \frac{1}{3!} \epsilon_{abc} e^a \wedge e^b \wedge e^c = e^{123}  , \qquad   *e_a = \frac{1}{2} \epsilon_{abc} e^{bc} ,\qquad         *e_{ab} = \epsilon_{abc} e^c ,\qquad *e_{abc} = \epsilon_{abc} .
      \end{align}
We fix the orientation of manifold by choosing $\epsilon_{123}=+1$. Further, we will need the following algebraic results
  \begin{align} \label{eq:epsilon_operations}
      \epsilon_{abc} \epsilon^{abc} = 3! , \quad \epsilon_{abc} \epsilon^{abm} = 2! \delta^m_c , \quad  \epsilon_{abc} \epsilon^{alm} = \delta^l_b \delta^m_c - \delta^l_c \delta^m_b ,\quad 
     \epsilon_{abc} \epsilon^{klm} = \begin{vmatrix}
          \delta^k_a & \delta^l_a &  \delta^m_a \\
          \delta^k_b & \delta^l_b &  \delta^m_b \\
          \delta^k_c & \delta^l_c &  \delta^m_c 
      \end{vmatrix}  .  
  \end{align}

 \subsection{Path independence of scalar product in the non-metric teleparallel geometries}

Scalar (inner) product of two vectors may be dependent of the path along which they are parallel propagated in non-metric geometries. In order to see this explicitly let us consider two vectors $U=U^a X_a$ and $V=V^a X_a$. Their scalar product is formulated by $(U,V)= \delta_{ab} U^a V^b$. Now, we parallel transfer it along a loop $\gamma$. Its total change is calculated as follows
 \begin{align}
     \Delta (U,V) =& \oint_\gamma d(U,V) = \oint_\gamma d\left( \delta_{ab} U^a V^b \right) \nonumber \\
      =& \oint_\gamma \big[  (D\delta_{ab}) U^aV^b +  \delta_{ab} (DU^a) V^b  +  \delta_{ab} U^a (DV^b)  \big] \nonumber \\
      =& -2\oint_\gamma Q_{ab}  U^aV^b + \oint_\gamma V_a (DU^a)   + \oint_\gamma  U_a (DV^a)  .
 \end{align}
On the other hand, parallel transport of vectors $U$ and $V$ along the closed curve $\gamma$ is formulated by $D_{\dot{\gamma}}U^a=0$ and $D_{\dot{\gamma}}V^a=0$ where $\dot{\gamma}$ denotes the tangent vector to the curve $\gamma$. They correspond to vanishing of the second and the third integrals. Thus, we arrive at
  \begin{align}
       \Delta (U,V)  = -2\oint_\gamma Q_{ab}  U^aV^b .
  \end{align}
Now, let $S$ be the surface bounded by the loop $\gamma$, i.e., $\partial S := \gamma$. Then, the usage of Stokes theorem, written as $\int_M d\Lambda = \int_{\partial M} \Lambda$ in the language of exterior forms where $\Lambda$ is any $p$-form, gives rise to 
   \begin{align}
        \Delta (U,V) =& -2\oint_{\partial S} Q_{ab}  U^aV^b  = -2\iint_S  d\left( Q_{ab} U^a V^b \right) \nonumber \\
           =& -2\iint_S (DQ_{ab})  U^aV^b + 2\iint_S Q_{ab} V^b (DU^a) + 2\iint_S Q_{ab} U^a (DV^b) \nonumber \\
           =&  -2\iint_S R_{(ab)}  U^aV^b.
   \end{align}
Again, the last two integrals are equal to zero in the second line because of the parallel transport rule of vectors. We used the first Bianchi identity (\ref{eq:bianchi1}) $DQ_{ab}=R_{(ab)}$ at the last step. Consequently, because of existence of non-metricity parallel transport of the scalar product seems depending on the path. But, in teleparallel geometries defined by zero-curvature, i.e.,  both  $R_{[ab]}=0$ and $R_{(ab)}=0$, it turns out to be independent of the path. This result is the same as the Proposition 2.1 of Ref.\cite{roychowdhury2017}. For more discussion on this issue one can consult \cite{adak2023sce} and the references therein.

\section{Review of defect theory in the metric affine geometry} \label{sec:defect-theory-present}


As a reference we firstly give brief information about the static defect theory written covariantly in terms of exterior algebra in the Riemann-Cartan-Weyl geometry. We mainly follow the results in \cite{povstenko1991},\cite{roychowdhury2017} in which $Q_{abc}$ is called as the density of metric anomalies. Density tensors of dislocation $\alpha^{ab}$, rotational disclination $\theta^{ab}$ and metrical disclination $\zeta_{ab}{}^{c}$ are related to torsion tensor $T^a{}_{bc}$, anti-symmetric components $R_{[ab]cd}$ and symmetric components $R_{(ab)cd}$ of curvature tensor $R^a{}_{bcd}$, respectively.
   \begin{subequations}
         \begin{align}
      \alpha^{ab} &= \frac{1}{2} \epsilon^{acd} T^b{}_{cd} , \label{eq:dislocation1}\\
       \theta^{ab} &= \frac{1}{4} \epsilon^{amn} \epsilon^{bkl} R_{[kl]mn} , \label{eq:rot-disclination1} \\
       \zeta_{ab}{}^c &= \frac{1}{2} \epsilon^{ckl} R_{(ab)kl} , \label{eq:met-disclination1}
  \end{align}
   \end{subequations}
where we did not insert explicitly the anti-symmetrization parentheses for $cd$ indices in the first relation, for $mn$ indices in the second one and for $kl$ in the third one because of appearance of the epsilon tensor. The metric disclination density tensor is symmetric at the first two indices by definition, $\zeta_{ab}{}^c=\zeta_{ba}{}^c$. Inverses of these relation could be obtained straightforwardly as 
    \begin{subequations}
         \begin{align}
     T^a{}_{bc} &= \epsilon_{bcd}  \alpha^{da} , \label{eq:dislocation1a}\\
      R_{[ab]cd}  &=  \epsilon_{abl} \epsilon_{cdk}  \theta^{kl} , \label{eq:rot-disclination1a} \\
       R_{(ab)cd}  &= \epsilon_{cdk} \zeta_{ab}{}^k  . \label{eq:met-disclination1a}
  \end{align}
   \end{subequations}
Physically, the diagonal and off-diagonal components of $\alpha^{ab}$ measure the density of screw and edge dislocations, respectively. Similarly the diagonal and off-diagonal components of $\theta^{ab}$ measure the density of wedge and twist disclinations, respectively \cite{dewit1973iii}. We could not find any information about the physical interpretation of the metrical disclination density tensor.

Now, we rewrite them in terms of exterior forms as 
  \begin{subequations} \label{eq:pairings1}
  \begin{align}
      \alpha^{ab} &= *(e^a \wedge T^b) , \label{eq:dislocation2} \\
      \theta^{ab} &= \frac{1}{2} \epsilon^{bcd} *\left(e^a \wedge R_{[cd]} \right) , \label{eq:rot-disclination2} \\
      \zeta_{ab}{}^c &= *\left(  R_{(ab)} \wedge e^c  \right) ,\label{eq:met-disclination2}
  \end{align}
  \end{subequations}
and
  \begin{subequations}
  \begin{align}
      T^a &= \alpha^{ba} *e_b , \label{eq:dislocation2a} \\
      R_{[ab]}  &=  \epsilon_{abc} \theta^{dc} *e_d , \label{eq:rot-disclination2a} \\
      R_{(ab)}   &=  \zeta_{ab}{}^c *e_c . \label{eq:met-disclination2a}
  \end{align}
  \end{subequations}
Then, the decomposition of the full curvature 2-form turns out to be
  \begin{align}
      R_{ab}= R_{[ab]} + R_{(ab)} = \epsilon_{abc} \theta^{dc} *e_d + \zeta_{ab}{}^c *e_c . \label{eq:disclination3}
  \end{align}
At this point it is worthwhile to remember the first Bianchi identity (\ref{eq:bianchi1}); $DQ_{ab}=R_{(ab)}$. Thus, in the teleparallel geometries which are defined by the condition $R^a{}_b =0$ meaning $R^{[ab]} \sim \theta^{ab} =0$ and $R_{(ab)} \sim \zeta_{ab}{}^c =0$, although we have non-metricity at our disposal, its covariant derivative and correspondingly both disclinations vanish.  

Covariant exterior derivative of (\ref{eq:dislocation2a}) yields the continuity condition for the dislocation density tensor. We will show it explicitly.
   \begin{subequations}
     \begin{align}
        D T^a &= D (\alpha^{ba} *e_b) \\
      R^a{}_b  \wedge e^b &= \left( D \alpha^{ba} \right) \wedge *e_b + \alpha^{ba} \left( D*e_b \right)  \\
      R^a{}_b  \wedge e^b &= \left( D_c \alpha^{ba} \right) e^c \wedge *e_b + \alpha^{ba} \left( -Q \wedge *e_b + *e_{bc} \wedge T^c \right)
      \end{align}
   \end{subequations}
Here we used the second Bianchi identity (\ref{eq:bianchi2}) on lhs and the algebraic identities (\ref{eq:identities-algebraic}) and (\ref{eq:identities-algebraic2}) on rhs. Now, by writing the non-metricity trace 1-form in components $Q=Q_c e^c = Q^a{}_{ac} e^c$ and substituting (\ref{eq:dislocation2a}) and (\ref{eq:disclination3}) here we arrive at the first continuity condition for $\alpha_{ab}$ as
  \begin{align}
      D_b\alpha^{ba} = \epsilon^{abc} \theta_{bc} + \epsilon_{bcd} \alpha^{cd} \alpha^{ba} + \zeta^{ab}{}_{b} + \alpha^{ba} Q_b . \label{eq:compatibilit1}
  \end{align}
Similarly, covariant exterior derivative of (\ref{eq:disclination3}) generates the second and third continuity conditions for the rotational and the metrical disclination tensors. In order to see it explicitly we start with the third Bianchi identity  (\ref{eq:bianchi3}) in the following form.
  \begin{align}
      DR^c{}_b=0 \quad \Rightarrow \quad  D\left(\delta^{ca} R_{ab} \right)=0  \quad \Rightarrow \quad  DR_{ab} = -2Q^c{}_a \wedge R_{cb}  \label{eq:bianci3a}
  \end{align}
The anti-symmetric part yields
  \begin{align}
        DR_{[ab]} = -Q^c{}_a \wedge R_{cb} + Q^c{}_b \wedge R_{ca} .
  \end{align}
We substitute (\ref{eq:rot-disclination2a}) into lhs and (\ref{eq:disclination3}) into rhs. After using the identities (\ref{eq:identities-algebraic}) and (\ref{eq:identities-algebraic2}) we insert (\ref{eq:dislocation2a}). Thus we arrive at the transient result.
  \begin{align}
      \epsilon_{abc} \left( D_m \theta^{mc} \right) =& \epsilon_{abc} \epsilon_{kmd} \alpha^{km} \theta^{dc} + 2 \epsilon_{abc} \theta^{dc} Q_d + \epsilon_{kbc} \theta^{dk} Q^c{}_{ad} + \epsilon_{akc} \theta^{dk} Q^c{}_{bd} \nonumber \\
      & - \zeta_{cb}{}^d  Q^c{}_{ad} + \zeta_{ca}{}^d  Q^c{}_{bd}  
  \end{align}
The second continuity condition for the rotational disclination tensor is obtained via multiplication of that by $\epsilon^{abi}$ and usage of (\ref{eq:epsilon_operations}) and the rearrangement of indices as
  \begin{align}
        D_b \theta^{ba}  = \epsilon_{bcd} \alpha^{bc} \theta^{da} + \theta^{ba} Q_b + \theta^{bc} Q^a{}_{cb} - \epsilon^{abc} \zeta_{mc}{}^k  Q^m{}_{bk} . \label{eq:compatibilit2}  
  \end{align}
Next, we calculate the symmetric part of (\ref{eq:bianci3a}).
  \begin{align}
        DR_{(ab)} = -Q^c{}_a \wedge R_{cb} - Q^c{}_b \wedge R_{ca} 
  \end{align}
Again, we substitute (\ref{eq:met-disclination2a}) into lhs and (\ref{eq:disclination3}) into rhs. After using the identities (\ref{eq:identities-algebraic}) and (\ref{eq:identities-algebraic2}) we insert  (\ref{eq:dislocation2a}). Thus we arrive at the third continuity condition for the metrical disclination tensor.
 \begin{align}
     D_c\zeta_{ab}{}^c = \epsilon_{ckd} \alpha^{kd} \zeta_{ab}{}^c + \zeta_{ab}{}^c Q_c + \theta^{mk} \epsilon_{ack}  Q^c{}_{bm} + \theta^{mk} \epsilon_{bck} Q^c{}_{am} - \zeta_{ac}{}^m Q^c{}_{bm} - \zeta_{bc}{}^m Q^c{}_{am} \label{eq:compatibilit3}
 \end{align}
Finally, we obtain the fourth continuity condition for the density of metric anomalies by performing a short calculation on the first Bianchi identity (\ref{eq:bianchi1}) with help of the equations  (\ref{eq:dislocation2a}) and (\ref{eq:met-disclination2a}).
  \begin{align}
      \epsilon^{abc} \left( D_a Q_{mnb}\right) = \zeta_{mn}{}^c - \alpha^{cp} Q_{mnp} \label{eq:compatibilit4}
  \end{align}

After obtaining the all continuity conditions we pass to a subcase called the semi-metric compatible connection which is characterized by
  \begin{align}
      Q_{ab} = \frac{1}{3} Q \delta_{ab}  \qquad \text{and} \qquad \zeta_{ab}{}^c = \frac{1}{3} \zeta^c \delta_{ab}
  \end{align}
where $\zeta^c = \zeta_{b}{}^{bc}$. Now, the four continuity conditions (\ref{eq:compatibilit1}), (\ref{eq:compatibilit2}),  (\ref{eq:compatibilit3}), (\ref{eq:compatibilit4}) are simplified, respectively,
  \begin{subequations} \label{eq:contunitiy-eqns-RCW}
      \begin{align}
            D_b\alpha^{ba} &= \epsilon^{abc} \theta_{bc} + \epsilon_{bcd} \alpha^{cd} \alpha^{ba} + \frac{1}{3}\zeta^a + \alpha^{ba} Q_b , \label{eq:compatibilit1a} \\
             D_b \theta^{ba}  &= \epsilon_{bcd} \alpha^{bc} \theta^{da} + \frac{4}{3}\theta^{ba} Q_b , \label{eq:compatibilit2a} \\
           D_a\zeta^a &= \epsilon_{abc} \alpha^{ab} \zeta^c + \zeta^a Q_a , \label{eq:compatibilit3a}\\
            \epsilon^{abc} \left( D_a Q_b \right) &= \zeta^c - \alpha^{cb} Q_b . \label{eq:compatibilit4a}
      \end{align}
  \end{subequations}
For the metric compatible connection defined by the conditions $Q_a=0$ and $\zeta^a=0$, we lands to the Riemann-Cartan geometry and the continuity equations turn out to be a much simpler forms 
   \begin{subequations} \label{eq:continuity-eqns-RC}
      \begin{align}
            D_b\alpha^{ba} &= \epsilon^{abc} \theta_{bc} + \epsilon_{bcd} \alpha^{cd} \alpha^{ba} , \label{eq:compatibilit1b} \\
             D_b \theta^{ba}  &= \epsilon_{bcd} \alpha^{bc} \theta^{da}  . \label{eq:compatibilit2b} 
      \end{align}
  \end{subequations}
These results are parallel to those of the Ref. \cite{Dereli1987}-\cite{dereli-vercin-1991}. Assuming $\alpha^{ab}$ and $\theta^{ab}$ to be small and of the same order, i.e. in the linear limit, we recover the well-known conservation laws or the kinematic equations of defects in a covariant formulation which are valid for even the non-Cartesian coordinates.
  \begin{subequations}
      \begin{align}
            \widetilde{D}_b\alpha^{ba} &= \epsilon^{abc} \theta_{bc}  , \label{eq:compatibilit1c} \\
             \widetilde{D}_b \theta^{ba}  &= 0 , \label{eq:compatibilit2c} 
      \end{align}
  \end{subequations}
where $\widetilde{D}$ denotes the covariant exterior derivative with respect to the Levi-Civita connection $\widetilde{\omega}^a{}_b$. The former means that if the disclination density is asymmetric, i.e. $\epsilon^{abc} \theta_{bc} \neq 0$, dislocations can only end on disclinations (within the body) and conversely must emerge from them, the latter says that disclinations cannot end inside the body.

In the Cartesian coordinates, $x^a=(x,y,z)$, the orthonormal coframe becomes $e^a = dx^a$ such that $de^a=0$ yielding $\widetilde{\omega}^a{}_b=0$ via the equation (\ref{eq:Levi-Civita}). Then, the linear equations of motion of defects are cast as follows
  \begin{align}
      \frac{\partial \alpha^{ba}}{\partial x^b} = \epsilon^{abc} \theta_{bc} , \qquad   \frac{\partial \theta^{ba}}{\partial x^b} = 0 .
  \end{align}
Accordingly, the characteristic vectors, the general Burgers vector $B_l$ and the Frank vector $\Omega_q$ are defined in terms of defect densities by de Wit in \cite{dewit1973ii} as
  \begin{subequations}
      \begin{align}
          B^l &= \iint_S (\alpha^{pl} + \epsilon^{l}{}_{qr} \theta^{pq} x^r) *e_p ,\\
          \Omega^q &= \iint_S \theta^{pq} *e_p . \label{eq:frank-vector2}
      \end{align}
  \end{subequations}
The second relation shows that the disclination density $\theta_{pq}$ represents the flux of Frank vector in the $x_q$ direction that crosses unit area of the surface $S$ normal to the $x_p$ direction. A similar interpretation can be recast for the first relation.

In summary, we obtained the four equations of motion  (\ref{eq:compatibilit1}), (\ref{eq:compatibilit2}),  (\ref{eq:compatibilit3}), (\ref{eq:compatibilit4}) for defects (dislocation plus disclination plus metric anomaly) in the Riemann-Cartan-Weyl (or metric affine) geometry defined by the triple $(Q_{ab},T^a,R^a{}_b)$. Non-metricity 1-form is defined as density of metric anomalies, torsion 2-form is related with dislocation, anti-symmetric and symmetric components of full curvature 2-form with rotational disclination and metrical disclination, respectively. Physical picture of dislocation and rotational disclination are clear in material science, but visualization of metrical disclination and metric anomaly is blurry. Therefore, we intend to offer a new formulation of defect theory by narrowing the geometric realm. 

\newpage

\section{A new formulation of defect theory in terms of the general teleparallel geometry} \label{sec:new-theory}

It is worthy to state explicitly that this section is the original content of our paper. We remember the decomposition of the full connection,  $\omega_{ab}=\widetilde{\omega}_{ab} +\mathrm{L}_{ab}$. Defect 1-form contains torsion (via contortion) and non-metricity, $\mathrm{L}_{ab} = K_{ab} + (\iota_bQ_{ac} -\iota_aQ_{bc})e^c + Q_{ab}$. Thus we realize that the anti-symmetric part is $\mathrm{L}_{[ab]} = K_{ab} + (\iota_bQ_{ac} -\iota_aQ_{bc})e^c := \Omega_{ab}$ and the symmetric part is $\mathrm{L}_{(ab)} = Q_{ab}$.

We postulate the following relation between disclination density tensor and the symmetric part of the defect 1-form, i.e, non-metricity 1-form,
 \begin{subequations} \label{eq:theta-nonmet}
 \begin{align}
     \theta^{ab} &= \frac{1}{3C} \epsilon^a{}_{cd} Q^{bcd} - \frac{1}{15C} \epsilon^{abc}Q_c , \label{eq:theta-from-nonmet1} \\
     Q^{bcd} &= C \left( \epsilon_k{}^{cd} \theta^{kb} + \epsilon_k{}^{bd} \theta^{kc}  +  \delta^{bc} \epsilon_{mn}{}^d \theta^{mn} \right) , \label{eq:nonmet-from-theta1} 
 \end{align}
 \end{subequations}
where $C$ is a constant which is kept as a suitable unit conversion parameter. They satisfy the properties $\delta_{ab}\theta^{ab}=0$ and $P^b =0$. The second equation can be rewritten as
 \begin{align}
     Q^{bc} = Q^{bcd} e_d = C \left( \theta^{kb} *e_k{}^c + \theta^{kc} *e_k{}^b  + \delta^{bc} \theta^{mn} *e_{mn} \right). \label{eq:discl2b}
 \end{align}

Similarly, we postulate the following relation between dislocation density tensor and the anti-symmetric part of the defect 1-form
   \begin{subequations} \label{eq:alpha-torsion}
     \begin{align}
     \alpha^{ab} &= *(\Omega^{bc} \wedge e_c{}^a) = *(e^a \wedge T^b) - *(Q^{bc} \wedge e_c{}^a) , \label{eq:disloc2a}\\
     T^a &= \left( \alpha^{ba} - 4C\theta^{ba} + C \theta^{ab} \right) *e_b  . \label{eq:disloc2b}
 \end{align}
 \end{subequations}
While passing to the second step we use the relation between contortion and torsion (\ref{eq:torsion-contor}) and the relevant identity in the equations (\ref{eq:identities-algebraic2}). We can compute
 \begin{align}
     \alpha^{ab} = \frac{1}{2} \epsilon^{amn} T^b{}_{mn} + 4C \theta^{ab} - C \theta^{ba} . \label{eq:disloc2bi}
 \end{align}
Thus, it is worthy to remark that when the non-metricity hence  disclination vanishes, the general dislocation density reduces to the (pure) dislocation density $\alpha^{ab} =  *(e^a \wedge T^b)$ and $T^a = \alpha^{ba} *e_b$ as claimed by de Wit \cite{dewit1970i}-\cite{dewit1973iv}. The next step is to obtain continuity equations for the defect tensors. Thus, the crucial point is to compute the covariant exterior derivative of the equations (\ref{eq:theta-from-nonmet1}) and (\ref{eq:disloc2bi}).

We will need the Bianchi identities in component form. Some details can be found in the Appendix.
  \begin{subequations}
      \begin{align}
        DQ^{abc} &= 2Q^a{}_k Q^{kbc} + 2Q^b{}_k Q^{kac} + 2Q^c{}_k Q^{abk} + \frac{1}{2} \iota^c \left( Q^{abk}  T_k -  R^{(ab)} \right) , \\
        DQ_{c} &=  2 Q^{ab} Q_{abc} + \frac{1}{2} \iota_c \left( Q_{d} T^d  - R \right), \\
        DT^a{}_{pk} &= \frac{1}{3}  \iota_p\iota_k \left( T^a{}_{bc} T^b \wedge e^c - R^a{}_b \wedge e^b \right) , \\
         D\epsilon^{abc} &= 2 Q^{da} \epsilon_d{}^{bc} + 2 Q^{dc} \epsilon_d{}^{ab} + 2 Q^{db} \epsilon_d{}^{ca}  - \epsilon^{abc} Q ,
      \end{align}
  \end{subequations}
where we denote $R= \delta_{ab} R^{ab}$. Besides, we want to remind again that we are in the general teleparallel geometry, i.e., $R^a{}_b=0$, $R_{(ab)}=0$, $R=0$ above. 

Firstly, we calculate the covariant exterior derivative of disclination tensor.
  \begin{subequations}
   \begin{align}
      D\theta^{ab} =& \frac{1}{3C} \left [ (D\epsilon^a{}_{cd})Q^{bcd}+\epsilon^a{}_{cd}(DQ^{bcd})\right]-\frac{1}{15C}\left [ (D\epsilon^{abc})Q_c + \epsilon^{abc}(DQ_c) \right] \\ 
         =&C\left ( \frac{76}{15}\theta^{ab}+\frac{4}{15}\theta^{ba} \right )\theta^{cd}*e_{cd} + \frac{1}{3}\alpha^{cd}\theta^{ab}*e_{cd} -\frac{1}{3}\alpha^{ca}\theta^{db}*e_{cd} \nonumber \\
      &+C\left ( \frac{41}{15}\theta^{ad}-\frac{2}{3}\theta^{da} \right )\theta^{cb} *e_{cd} +C\left ( \frac{4}{15}\theta^{bd}+ \frac{4}{3}\theta^{db} \right )\theta^{ca}*e_{cd} \nonumber \\
      &+\frac{1}{6}\left ( \alpha^{cf}-4C\theta^{cf}+C\theta^{fc}\delta^{ab}\theta^d{}_f*e_{cd} \right ) + C\left ( \frac{46}{15}\theta^{ad} -\frac{2}{3}\theta^{da} \right )\theta_{cd}*e^{cb} \nonumber \\
      &-\frac{1}{6}\alpha_{cd}\theta^{ad}*e^{cb}-\frac{C}{6}\theta_{dc}\theta^{ad}*e^{cb} + \frac{4C}{15}\theta^{bd}\theta_{cd}*e^{ca}
  \end{align}
   \end{subequations}
Secondly, we calculate the covariant exterior derivative of dislocation tensor.
   \begin{subequations}
      \begin{align}
    D\alpha^{ab}=&\frac{1}{2}\left ( D\epsilon^{amn} \right )T^b{}_{mn} +\frac{1}{2}\epsilon^{amn}\left ( DT^b{}_{mn} \right ) + 4CD\theta^{ab}-CD\theta^{ba} \\
     =&C\left (5\alpha^{ab} +\theta^{ba} \right )\theta^{cd}*e_{cd} +\frac{1}{3}\left ( \alpha^{db}-4C\theta^{db}+C\theta^{bd} \right ) \left ( \alpha^{cf}-5C\theta^{cf} \right )\epsilon_{cfd}e^a \nonumber \\
     &+\frac{C}{3}\alpha^{cd}\left ( 4\theta^{ab}-\theta^{ba} \right )*e_{cd} +C^2\left ( \frac{32}{3}\theta^{ad}-4\theta^{da} \right )\theta^{cb}*e_{cd} \nonumber \\
     &+C^2\left ( 12\theta^{ad}-\frac{8}{3}\theta^{da} \right )\theta_{cd}*e^{cb} + C^2\left ( \frac{2}{3}\theta^{db}-2\theta^{bd} \right )\theta_{cd}*e^{ca} \nonumber \\
     &+C^2\left ( 6\theta^{db}-\frac{5}{3}\theta^{bd} \right )\theta^{ca}*e_{cd}+\frac{1}{2}\left ( \alpha^{cf}-4C\theta^{cf} +C\theta^{fc} \right )\theta^d{}_d\delta^{ab}*e_{cf}\nonumber \\
     &+C\alpha_{cd}\left ( \theta^{bd}*e^{ca}-\theta^{ad}*e^{cb} \right ) +C^2\theta_{dc}\left ( \frac{1}{6}\theta^{bd}*e^{ca}-\frac{2}{3}\theta^{ad}*e^{cb} \right ) \nonumber \\
     &+C\left ( \frac{1}{3}\alpha^{cb}\theta^{da}-\frac{4}{3}\alpha^{ca}\theta^{db} \right )*e_{cd}
   \end{align}
 \end{subequations}
Thus we compute the covariant divergence of defect tensors by hitting $\iota_a$ to the found results. 
  \begin{subequations}
  \begin{align}
     D_a\theta^{ab}=&\frac{C}{3}\theta^{ab}\theta^{cd}\epsilon_{cda}+\frac{2}{3}\alpha^{cd}\theta^{ab}\epsilon_{cda} + C\left ( \frac{46}{15}\theta^{ad}-\frac{2}{3}\theta^{da} \right )\theta_{cd}\epsilon^{cb}{}_a \nonumber \\
     &+C\left ( \frac{2}{3}\theta_{cd}-\frac{1}{3}\theta_{dc} \right )\theta^{ad}\epsilon^{cb}{}_a -\frac{1}{3}\alpha_{cd}\theta^{ad}\epsilon^{cb}{}_a \\
          D_a\alpha^{ab}=&\frac{1}{2}\alpha^{ab}\alpha^{cd}\epsilon_{cda}+\frac{11C}{3}\alpha^{ab}\theta^{cd}\epsilon_{cda}+\frac{4C}{3}\alpha^{cd}\theta^{ab}\epsilon_{cda}+C^2\left ( \theta^{ba}-\theta^{ab} \right )\theta^{cd}\epsilon_{cda} \nonumber \\
     &+C^2\left ( 10\theta^{ad}-\frac{13}{6}\theta^{da} \right )\theta_{cd}\epsilon^{cb}{}_a-\left (\frac{3C}{2}\alpha_{cd}+\frac{2C^2}{3}\theta_{dc}\right )\theta^{ad}\epsilon^{cb}{}_a
 \end{align}
  \end{subequations}
These are our continuity equations such as those (\ref{eq:continuity-eqns-RC}). We want again to emphasize  that although the Riemann-Cartan geometry in which standard geometric formulation of defect theory is established does not have a metric formulation, the general teleparallel geometry does have a metric formulation \cite{adak2023ijgmmp}. This result is one of motivations directing us to do this work.

\section{Discussion}

Since dislocation density and disclination density are useful quantities to obtain information about many elastic fields such as distortion, strain, dilatation, stress and rotation caused by defects in a continuum, it makes sense to look for a complete theory that explains their behavior. One of the various ways to do that is to adopt the geometric approach. As far as we know, in almost all studies conducted in this direction in the literature, the dislocation density is matched with the torsion tensor of the Riemann-Cartan geometry, and the disclination density with the full curvature tensor \cite{povstenko1991}-\cite{katanaev2021}. For a summary of that formulation, one can look at the equations (\ref{eq:continuity-eqns-RC}) and later of Section \ref{sec:defect-theory-present}. Those arguments are based on the definitions given by equations (\ref{eq:pairings1}), together with $Q_{ab}=0$, hence $R_{(ab)}=0$ and $\zeta_{abc}=0$ constraints. In that formulation, the geometric arena suitable for discussing the behaviour of only dislocation type defects appears naturally as metric (Weitzenböck) teleparallel geometry. Conversely, we are left with Riemannian geometry as the appropriate geometry for dislocation-free continuum.

Looking at that formulation from a different perspective will give us new insights. If one inserts the decomposition of full curvature given by the equation (\ref{eq:curv-ant-symm1}) or (\ref{eq:decompos-curv}), together with the metricity condition, into the definition  (\ref{eq:rot-disclination2}), it becomes clear that there are contributions to the disclination density from the torsion tensor via the contortion 1-form inside of $\mathrm{L}_{ab}$. Therefore, according to the pairing between torsion and dislocation density (\ref{eq:dislocation2}), since they are proportional to each other, there are contributions coming from the dislocation density to the disclination density. However, this conclusion is the opposite of what de Wit says in the papers \cite{dewit1970i}-\cite{dewit1973iv}. As we stated in the introduction, we adopt de Wit's terminology and approach regarding the continuum defects. According to de Wit's articles, if there is dislocation in a continuum, it contains some of them in the (general) dislocation density. When disclination is eliminated, the general dislocation density turns into (pure) dislocation density. To achieve this phenomenological result, we matched the symmetric part of the defect 1-form with the disclination density, see (\ref{eq:theta-nonmet}).  Accordingly, we also matched the dislocation density to the anti-symmetric part of the defect 1-form, see (\ref{eq:alpha-torsion}). These correspondences appear the first time in this paper. Thus, a disclination-free continuum, $\theta_{ab}=0$ hence $Q_{ab}=0$, can be represented by again the metric (Weitzenböck) teleparallel geometry. The reverse case characterized by the configuration, $T^a=0$ hence $\alpha_{ab}=0$, but $\theta_{ab} \neq 0$, is not always possible because of the equation (\ref{eq:disloc2a}). This result is in accordance with de Wit's conclusion. The geometric realm of the latter case is the symmetric teleparallel geometry.

It is worthy to notice that one way to satisfy mathematically de Wit's conclusion in the Riemann-Cartan geometry is to map dislocation to full curvature 2-form and disclination to torsion 2-form, opposite to the current approach in the literature.
However, working in all three teleparallel geometries has some advantageous compared with the Riemann-Cartan geometry. It is known that the metric (hence the orthonormal 1-form $e^a$) and the full connection 1-form $\omega^a{}_b$ are, in general, independent quantities in metric affine geometry. Nevertheless, although there is not any metric formulation of the Riemann-Cartan geometry, we have shown a way for metric formulation of the general teleparallel geometry in our recent paper \cite{adak2023ijgmmp} in which previously known metric formulations of the symmetric teleparallel geometry and Weitzenböck teleparallel geometry have been summarized as well.

The ground-state configurations of a continuum with defects can be determined by the extrema of a free-energy integral $F=\int_B L$ where $B$ is the body manifold corresponding to the three dimensional abstract manifold $M$ and $L$ is called the Lagrange 3-form   \cite{Dereli1987}-\cite{lazar-hehl-2010}. One of our future projects is to write down some possible Lagrange 3-forms containing torsion and non-metricity, obtain variational field equations, find some classes of exact solutions and compare them with the results coming from the material science world. Thus, we hope that our new and strict mathematical model for material defects may predict new insights on the behavior of material. 

In this work we introduced a new formulation of defect fields for static configurations. A natural extension of that is to consider its dynamical counterpart. By inclusion of the defect velocity tensor field similar to those in  \cite{azab-po-2020},\cite{schimming-2023} the results of our work could be generalized. That generalization remains an open problem waiting to be solved.

Our other future project is to apply our findings and experiences gained in this paper to studies of photonic crystals. A photonic crystal has a lattice pattern formed basically by using two different optical materials. If they are placed periodically in two dimensions, that structure is called a two-dimensional photonic crystal. Of course, there may be three-dimensional photonic crystals. In practice the two-dimensional ones which are constructed by arranging very tiny (in nanometer scale) dielectric roads at lattice points are the most commonly investigated. The aim of these searches is mainly to control the behaviour of electromagnetic wave by creating crystal defects in lattice pattern. Defects can be produced by various methods such as by changing the radius or the dielectric constant, by removing a selected rod, by adding a tiny auxiliary extra rod to some main roads etc. \cite{adak-photonic-2023i},\cite{adak-photonic-2023ii} and the references therein.  Correspondingly, a medium containing defects could be represented by a suitable teleparallel geometry and the Maxwell field could be coupled to it in a scale invariant way. Scale invariance is crucial because, in engineering of photonic crystal apparatus,  firstly some simulations are done, then experiments are performed in millimeter scale and finally fabrication process takes place in nanometer scale.

\appendix

\section*{Appendix}

\section{Bianchi identities}

We calculate some more identities from the Bianchi identities (\ref{eq:bianchi_identities}). We start with the first Bianchi identity (\ref{eq:bianchi1}). 
  \begin{align}
      D(Q_{abc} e^c)=R_{(ab)} \qquad \Rightarrow \qquad 
      (DQ_{abc}) \wedge e^c = -Q_{abc} T^c + R_{(ab)} .
  \end{align}
We hit firstly $\iota_k$ and then $\iota_m$, 
   \begin{subequations}
  \begin{align}
       (D_kQ_{abc}) \wedge e^c - DQ_{abk} &= \iota_k \left(  -Q_{abc}  T^c +  R_{(ab)} \right),\\
       D_kQ_{abm} - D_mQ_{abk} &=  \frac{1}{2} (\iota_m\iota_k - \iota_k\iota_m) \left(  -Q_{abc}  T^c +  R_{(ab)} \right)  ,
  \end{align}
   \end{subequations}
where we anti-symmetrized $mk$ indices on the right hand side of the second line. We conclude 
  \begin{align}
      D_kQ_{abm} = \frac{1}{2} \iota_k\iota_m \left( Q_{abc}  T^c - R_{(ab)} \right) \qquad \Rightarrow \qquad  DQ_{abm} = \frac{1}{2} \iota_m\left( Q_{abc}  T^c - R_{(ab)} \right) .
  \end{align}
Here it is not difficult to see $ DQ_{abm} \wedge *e^m = D^mQ_{abm} *1 =0$. Now we can compute 
 \begin{align}
     D*Q_{ab} =& D\left( Q_{abc} *e^c \right) = \left( DQ_{abc}\right) \wedge *e^c + Q_{abc} \left(  D*e^c \right) = Q_{abc} \left(  D*e^c \right) \nonumber \\
     =& Q_{abc} D (\delta^{mc} *e_m) = Q_{abc} \left(  2 Q^{mc} \wedge *e_m - Q \wedge *e^c + *e^c{}_d \wedge T^d\right) \nonumber \\
     =& (2 P  - Q -T) \wedge *Q_{ab}
 \end{align}
where $Q=\delta_{ab} Q^{ab}$, $P=(\iota^b Q_{ab}) e^a$ and $T=\iota_a T^a$. By the usage of $D\delta^a_b=0$, $D\delta_{ab}=-2Q_{ab}$ and $D\delta^{ab}=+2Q_{ab}$ this conclusion turns out to be
  \begin{align}
     D*Q^{ab} = (2 P  - Q -T) \wedge *Q^{ab} + 4 Q^a{}_c \wedge *Q^{cb} .
 \end{align}
 
Next, we start with the second Bianchi identity (\ref{eq:bianchi2}).
  \begin{align}
     \frac{1}{2} D(T^a{}_{bc} e^{bc}) = R^a{}_b \wedge e^b \qquad \Rightarrow \qquad (D T^a{}_{bc}) \wedge e^{bc} =  2 X^a
  \end{align}
where we defined a new three-form $X^a :=  R^a{}_b \wedge e^b  -T^a{}_{bc} T^b \wedge e^c$ for simplicity. Then we hit $\iota_m$ and $\iota_p$ and $\iota_k$ consecutively, and rearrange the terms
 \begin{align}
     D_mT^a{}_{pk} + D_pT^a{}_{km} + D_kT^a{}_{mp} = - \frac{1}{3} (\iota_m \iota_p \iota_k + \iota_p \iota_k \iota_m + \iota_k \iota_m \iota_p) X^a .
 \end{align}
Here since $mpk$ indices are cyclic on lhs, we make them cyclic on rhs. Thus, we conclude
  \begin{align}
      D_mT^a{}_{pk} = -\frac{1}{3} \iota_m \iota_p\iota_k X^a \qquad \Rightarrow \qquad DT^a{}_{pk} = -\frac{1}{3}  \iota_p\iota_k X^a .
  \end{align}
When one remembers that any four-form, e.g. $X^a \wedge *e^{pk}$, vanishes in three dimensions, it is trivial to see $DT^a{}_{pk} \wedge *e^{pk} =0$. Consequently, we compute
  \begin{align}
      D*T^a =& \frac{1}{2} D(T^a{}_{bc} *e^{bc}) = \frac{1}{2} (DT^a{}_{bc}) \wedge *e^{bc} + \frac{1}{2} T^a{}_{bc} (D*e^{bc} )= \frac{1}{2} T^a{}_{bc} (D*e^{bc})  \nonumber \\
      =&  \frac{1}{2} T^a{}_{bc} [  (D\delta^{bk}) \delta^{cm}*e_{km} +  \delta^{bk} (D\delta^{cm}) *e_{km} + \delta^{bk} \delta^{cm} (D*e_{km}) ] \nonumber \\
      =& 2 Q_{bc} \wedge *( e^c \wedge \iota^b T^a) - Q \wedge *T^a + T^b \wedge *(e_b \wedge T^a)      
  \end{align}

In spite of teleparallelism, $R^a{}_b=0$ in the paper, for the appendix to become complete on its own we perform the similar computations for the third Bianchi identity (\ref{eq:bianchi3}). 
  \begin{align}
     \frac{1}{2} D(R^a{}_{bcd} e^{cd}) = 0 \qquad \Rightarrow \qquad (D R^a{}_{bcd}) \wedge e^{cd} =  - 2 Z^a{}_b
  \end{align}
where we defined a new three-form $Z^a{}_b :=  R^a{}_{bcd} T^c \wedge e^d$ for simplicity. Then we hit $\iota_m$ and $\iota_p$ and $\iota_k$ consecutively, and rearrange the terms
 \begin{align}
     D_mR^a{}_{bpk} + D_pR^a{}_{bkm} + D_kR^a{}_{bmp} = \frac{1}{3} (\iota_m \iota_p \iota_k + \iota_p \iota_k \iota_m + \iota_k \iota_m \iota_p) Z^a{}_b
 \end{align}
where we made cyclic $mpk$ indices on the right hand side. Thus, we conclude
  \begin{align}
      D_mR^a{}_{bpk} = \frac{1}{3} \iota_m \iota_p\iota_k Z^a{}_b \qquad \Rightarrow \qquad DR^a{}_{bpk} = \frac{1}{3}  \iota_p\iota_k Z^a{}_b .
  \end{align}
When one remembers that any four-form, e.g. $Z^a{}_b \wedge *e^{pk}$, vanishes in three dimensions, it is trivial to see $DR^a{}_{bpk} \wedge *e^{pk} =0$. Consequently, we compute
  \begin{align}
      D*R^a{}_b =& \frac{1}{2} D(R^a{}_{bcd} *e^{cd}) = \frac{1}{2} (DR^a{}_{bcd}) \wedge *e^{cd} + \frac{1}{2} R^a{}_{bcd} (D*e^{cd} ) = \frac{1}{2} R^a{}_{bcd} (D*e^{cd})  \nonumber \\
      =&  \frac{1}{2} R^a{}_{bcd} [  (D\delta^{ck}) \delta^{dm}*e_{km} +  \delta^{ck} (D\delta^{dm}) *e_{km} + \delta^{ck} \delta^{dm} (D*e_{km}) ] \nonumber \\
      =& 2 Q_{cf} \wedge *( e^f \wedge \iota^c R^a{}_b) - Q \wedge *R^a{}_b + T^c \wedge *(e_c \wedge R^a{}_b) .     
  \end{align}

\section{Relation between disclination and non-metricity}

We postulate the following relations between disclination density tensor and non-metricity tensor
 \begin{subequations}
 \begin{align}
     \theta^{ab} &= A \epsilon^a{}_{cd} Q^{bcd} + B \epsilon^{abc}Q_c , \label{eq:theta-from-nonmet} \\
     Q^{bcd} &= C \left( \epsilon_k{}^{cd} \theta^{kb} + \epsilon_k{}^{bd} \theta^{kc} \right) + K \delta^{bc} \epsilon^d{}_{mn} \theta^{mn} , \label{eq:nonmet-from-theta} 
 \end{align}
 \end{subequations}
where $A,B,C,K$ are arbitrary constants. The first relation satisfies the property $\delta_{ab} \theta^{ab}=0$. We denote $Q_{abc} = \iota_cQ_{ab}$ and $Q_c=\iota_cQ = \delta^{ab} Q_{abc} = Q^a{}_{ac}$ where $Q_{(ab)c}$ by definition. We also define the second trace $P_a = \delta^{bc} Q_{abc} = \iota^b Q_{ab} = Q_{ab}{}^{b}$ along with the first trace $Q_c = \delta^{ab} Q_{abc}$ of non-metricity. We compute the second kind trace from the equation (\ref{eq:nonmet-from-theta}) as  
 \begin{align}
     P^b = \delta_{cd} Q^{bcd} = (K-C) \epsilon^b{}_{mn} \theta^{mn} .
 \end{align}
We prefer to set $P^a=0$ from the outset for mathematical simplification by choosing $K=C$. Then we compute the first trace as
 \begin{align}
    Q^d = \delta_{bc} Q^{bcd} = 5C \epsilon^d{}_{mn} \theta^{mn} .
 \end{align}
As seen, choice of $P^b=0$ for mathematical reasoning does not cause any loss of physical generality. Now by substituting (\ref{eq:nonmet-from-theta}) into (\ref{eq:theta-from-nonmet}) we obtain
 \begin{align}
     \theta^{ab} &= AC \epsilon^a{}_{cd} \left( \epsilon_k{}^{cd} \theta^{kb} + \epsilon_k{}^{bd} \theta^{kc} + \delta^{bc} \epsilon^d{}_{mn} \theta^{mn} \right) + 5 BC \epsilon^{abc} \epsilon_{mnc} \theta^{mn} \nonumber \\
     &= (4AC +5BC) \theta^{ab} - (AC + 5BC) \theta^{bc} .
 \end{align}
Symmetric part and anti-symmetric parts yield two relations, respectively,
 \begin{align}
     3AC = 1 \quad \text{and} \quad 5AC + 10 BC = 1 .
 \end{align}
Thus, we conclude
 \begin{align}
     A= \frac{1}{3C} \quad \text{and} \quad B= - \frac{1}{15C} .
 \end{align}
Reversely, we insert (\ref{eq:theta-from-nonmet}) into   (\ref{eq:nonmet-from-theta}) 
 \begin{align}
     Q^{bcd} &= C \epsilon_k{}^{cd} \left( A \epsilon^{kaf} Q^{b}{}_{af} + B \epsilon^{kba}Q_a \right) +  C \epsilon_k{}^{bd} \left( A \epsilon^{kaf} Q^{c}{}_{af} + B \epsilon^{kca}Q_a \right) \nonumber \\
     & \quad +  C \delta^{bc} \epsilon^d{}_{mn} \left( A \epsilon^{maf} Q^{n}{}_{af} + B \epsilon^{mna}Q_a \right) \nonumber \\
     &= AC \left( 2Q^{bcd} - Q^{dbc} - Q^{dcb} + \delta^{bc} Q^d \right) + BC \left( 4 \delta^{bc} Q^d - \delta^{db} Q^c - \delta^{dc} Q^b \right)
 \end{align}
By hitting $\delta_{bc}$ we arrive at the same condition $5AC + 10 BC = 1$ above. By multiplying with $\delta_{cd}$ we check the property $P^b=0$. Consequently we leave the constant $C$ as a suitable unit conversion parameter. 




\noindent 
{\bf Acknowledgements:} TD acknowledges partial support by the Turkish Academy of Sciences.

 
 \noindent
{\bf Data Availability Statement:} No data associated in the manuscript.

\end{document}